\documentstyle[psfig,aps,prl]{revtex}
\begin{document}
\title{Bayesian Thermostatistical Analyses of Two-Level Complex and
Quaternionic Quantum Systems}
\author{Paul B. Slater}
\address{ISBER, University of
California, Santa Barbara, CA 93106-2150\\
e-mail: slater@itp.ucsb.edu,
FAX: (805) 893-7995}

\date{\today}

\draft
\maketitle
\vskip -0.1cm

\begin{abstract}
The three and five-dimensional convex sets of two-level complex and
quaternionic quantum systems are studied in the Bayesian thermostatistical
framework introduced by Lavenda. Associated with a given parameterization of
each such set is a quantum Fisher (Helstrom) information matrix. The square
root of its determinant --- adopting an {\it ansatz} of Harold Jeffreys --- provides a reparameterization-invariant prior measure over the set. Both such
measures can be properly normalized and their univariate marginal probability
distributions --- which serve as structure functions --- obtained.
Gibbs (posterior) probability distributions can then be found, using Poisson's
integral representation of the modified spherical Bessel functions.
The square roots of the (classical) Fisher information of these Gibbs
distributions yield (unnormalized) priors over the inverse temperature
parameters.
\end{abstract}


\vspace{-0.1cm}

In this letter, we apply the Bayesian thermostatistical framework of Lavenda
\cite{lav1,lav2,lav3} to the two-level quantum systems, both complex and
quaternionic \cite{adler} in nature. A two-level quaternionic system
can be represented by a $2 \times 2$ density matrix
\begin{equation} \label{1}
\rho = {1 \over 2} \pmatrix{1+z &x-iy-ju-kv\cr
x+iy +ju +kv & 1-z\cr},
\end{equation}
where $i^2 =j^2 =k^2 =-1,ij=-ji=k, jk=-kj=i$ and $ki =-ik=j$.
Setting $u=v=0$, we obtain the familiar (Pauli matrix) representation of the
two-level complex systems \cite[sec. 4.2]{belt}, in which the points
$(x,y,z)$ lie within the unit ball ($x^2 +y^2 +z^2 < 1$).

In the complex case ($u=v=0$), one can --- using the concept of a symmetrized
logarithmic derivative \cite{hels,khol,mall,fuji,brau,slat} --- associate with
the given parameterization, the quantum Fisher information matrix,
\begin{equation} \label{2}
{1 \over (1-x^2-y^2-z^2)} \pmatrix{1-y^2-z^2 & xy & xz\cr
  xy & 1-x^2-z^2 & yz \cr
xz&yz &1-x^2-y^2\cr}.
\end{equation}
In the quaternionic instance, employing the relations between the Pauli
matrices and the quaternions \cite[p. 495]{adler} \cite[p. 197]{bied} to
generate a complex $4 \times 4$ complex density matrix and then finding the
corresponding symmetrized logarithmic derivatives \cite{slat} --- one has
\begin{equation} \label{3}
 {1 \over (1-u^2 -v^2 -x^2-y^2 -z^2)} \pmatrix{ g(\{u\}) & uv & ux &
uy & uz \cr
uv & g(\{v\}) & vx &vy & vz \cr
ux & vx & g(\{x\}) & xy & xz \cr
uy & vy & xy & g(\{y\}) & yz \cr
uz & vz & xz & yz & g(\{z\}) \cr},
\end{equation}
where $g(\{u\})$ denotes $1-v^2-x^2-y^2-z^2$, $\ldots$
(The inverses of (\ref{2}) and (\ref{3}) provide --- by the Cram\'er-Rao 
inequality --- lower bounds [in the sense of negative definiteness] on the
covariance matrices of the parameters \cite{hels,khol}.) The determinants
of (\ref{2}) and (\ref{3}) are, respectively,
\begin{equation} \label{4}
{1 \over (1-x^2-y^2-z^2)}
\end{equation}
and
\begin{equation} \label{5}
{1 \over (1-u^2-v^2-x^2-y^2-z^2)}.
\end{equation}
(These are inversely proportional to the determinants obtained from
(\ref{1}) itself.)

In the classical/nonquantum situation --- in which probability distributions
rather than density matrices are studied --- Harold
 Jeffreys proposed that one employ
the square root of the determinant of a Fisher information matrix as a
(reparameterization-invariant) prior measure (volume element) over a family
(Riemannian manifold) of probability distributions \cite{jeff1,jeff2,bern}.
Adopting this ansatz to the quantum mechanical case, one 
can --- normalizing the square roots of (\ref{4}) and (\ref{5}) over the
three and five-dimensional unit balls --- derive the prior probability
distributions,
\begin{equation} \label{6}
{1 \over \pi^2 (1-x^2-y^2-z^2)^{1/2}}, \qquad (x^2+y^2+z^2 \leq 1)
\end{equation}
and
\begin{equation} \label{7}
{2 \over \pi^3 (1-u^2-v^2-x^2-y^2-z^2)^{1/2}}. \qquad
(u^2 +v^2 +x^2 +y^2 +z^2 \leq 1)
\end{equation}
The three bivariate marginal probability distributions of (\ref{6}) are
uniform over unit disks, while the five quadrivariate marginal probability
distributions of (\ref{7}) are uniform over unit balls in four-space.
Further integrating over all but the last
 coordinate, one arrives at a {\it univariate}
probability distribution, having the form, in the complex case,
\begin{equation} \label{8}
{2 (1-z^2)^{1/2} \over \pi}, \qquad (-1 \leq z \leq 1)
\end{equation}
and, in the quaternionic instance,
\begin{equation} \label{9}
{8 (1-z^2)^{3/2} \over 3 \pi}. \qquad (-1 \leq z \leq 1).
\end{equation}
We consider (\ref{8}) and (\ref{9}) to be (normalized) structure functions
in the sense of Lavenda \cite{lav1,lav2,lav3}.

\begin{figure}
\centerline{\psfig{figure=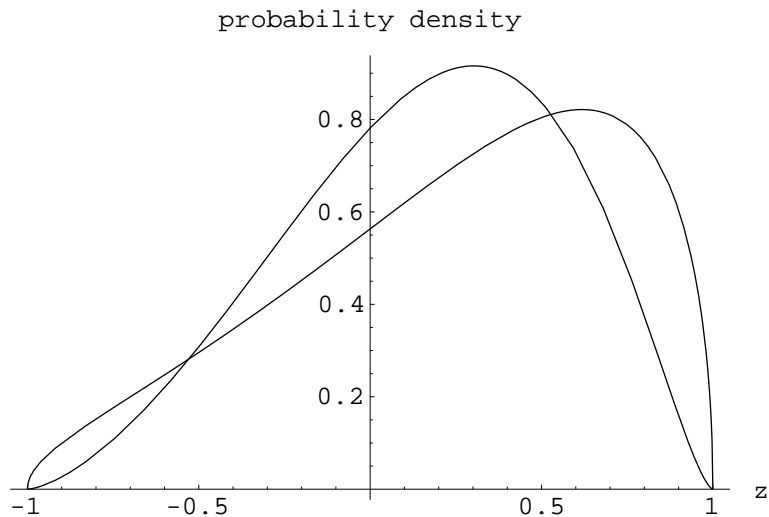}}
\caption{Gibbs distributions (\ref{12}) for $\beta = -1$. The higher-peaked
curve corresponds to the quaternionic ($n=2$) case.}
\label{f1}
\end{figure}

\begin{figure}
\centerline{\psfig{figure=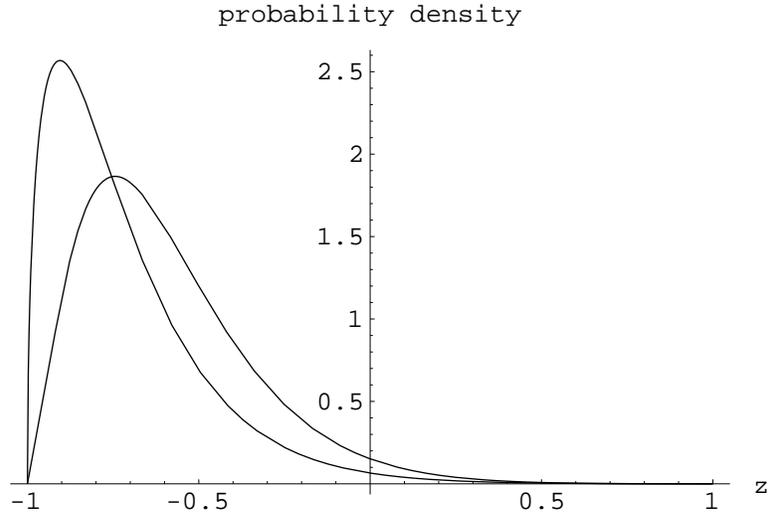}}
\caption{Gibbs distributions (\ref{12}) for $\beta = 5$. The lower-peaked curve
corresponds to the quaternionic ($n=2$) case.}
\label{f2}
\end{figure}

Since (\ref{8}) and (\ref{9}) are proportional to expressions of the type,
\begin{equation} \label{10}
(1-z^2)^{n-1/2}, \qquad (n=1,2)
\end{equation}
one can employ Poisson's integral representation of the modified spherical
Bessel functions (cylinder functions of half integral order) \cite{karm},
\begin{equation} \label{11}
I_{n}(\beta) = {({\beta \over 2})^{n} \over \sqrt{\pi}
 \Gamma (n + {1 \over 2})}
\int_{-1}^{1} \mbox{exp} (-\beta z) (1-z^2)^{n-1/2} \mbox{d} z
\end{equation}
to introduce thermodynamic considerations \cite{park,band,slat2}. One then has
(properly normalized) Gibbs distributions of the form,
\begin{equation} \label{12}
{\mbox{exp} (-\beta z) ({\beta \over 2})^{n} (1-z^2)^{n-1/2} \over
I_{n}(\beta) \sqrt{\pi} \Gamma (n + {1 \over 2})},
\end{equation}
where $\beta$ serves as the inverse temperature parameter. For $\beta = -1$,
Fig.~\ref{f1} shows these distributions for $n=1$ (complex) and $n=2$
(quaternionic), the latter curve having the higher peak.
 Fig.~\ref{f2}
displays the analogous results for $\beta=5$, with the curve for the
complex case now having the higher peak.

\begin{figure}
\centerline{\psfig{figure=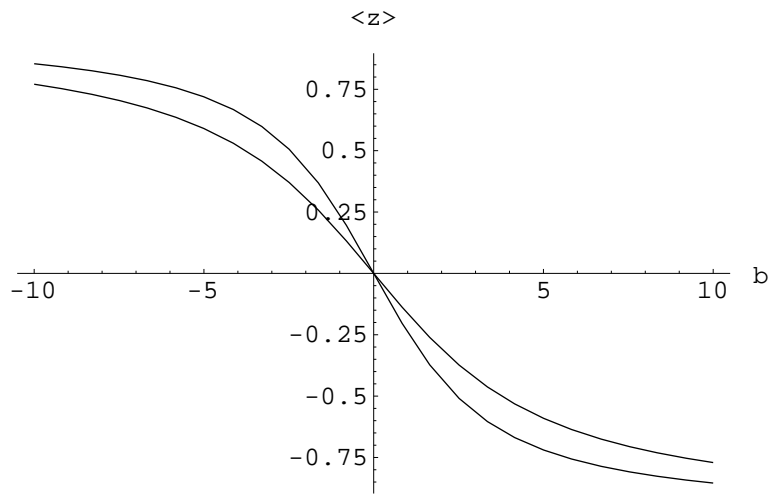}}
\caption{Expected value of $z$ as a function of $\beta$. The quaternionic
($n=2$)  curve is the flatter one.}
\label{f3}
\end{figure}

\begin{figure}
\centerline{\psfig{figure=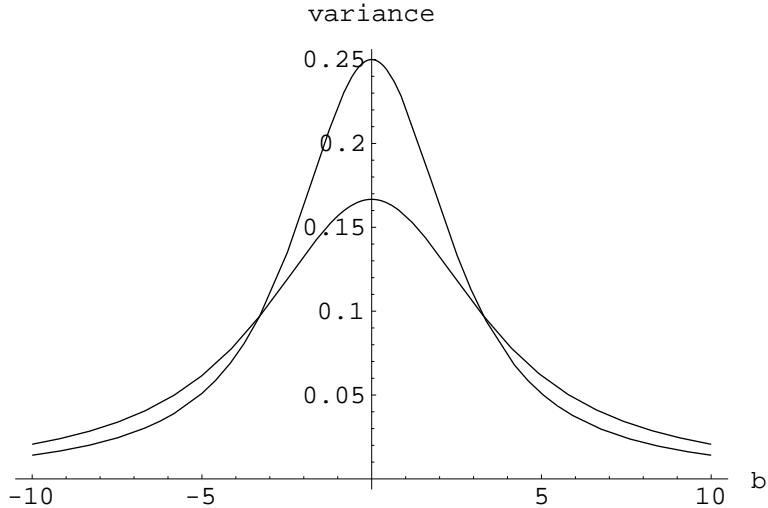}}
\caption{The variance about  the
 expected value of $z$ as a function of $\beta$.
The quaternionic ($n=2$) curve is the flatter one.}
\label{f4}
\end{figure}

\begin{figure}
\centerline{\psfig{figure=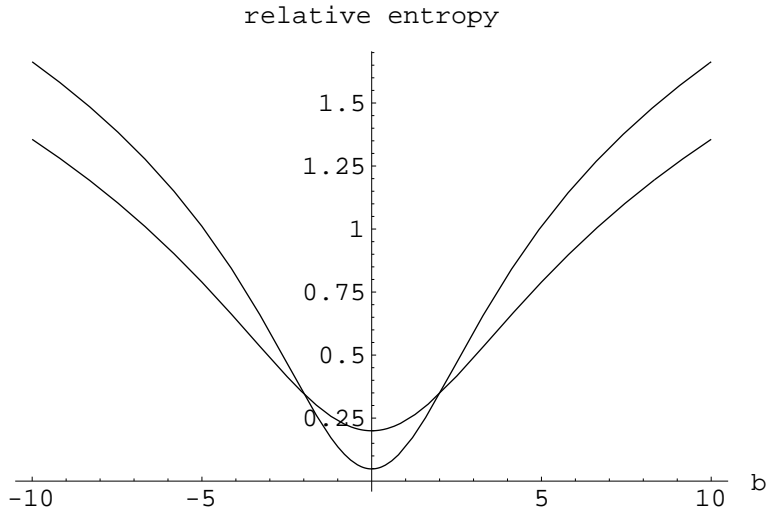}}
\caption{Relative entropy of Gibbs distributions with respect to a uniform
distribution. The quaternionic ($n=2$) curve is the one having
 a greater minimum.}
\label{f5}
\end{figure}

In Fig.~\ref{f3} and \ref{f4} are shown, respectively, the expected value of
$z$  ($<z>$) and the variance about $<z>$
 as a function of $\beta$, with the quaternionic
curves being the flatter ones in the two graphs.
 In Fig.~\ref{f5} are
displayed the relative entropies of the Gibbs distributions (\ref{12}) with
respect to the uniform distribution (${1 \over 2}$) over $z \in [-1,1]$, with
the quaternionic curve having the greater minimum.

\begin{figure}
\centerline{\psfig{figure=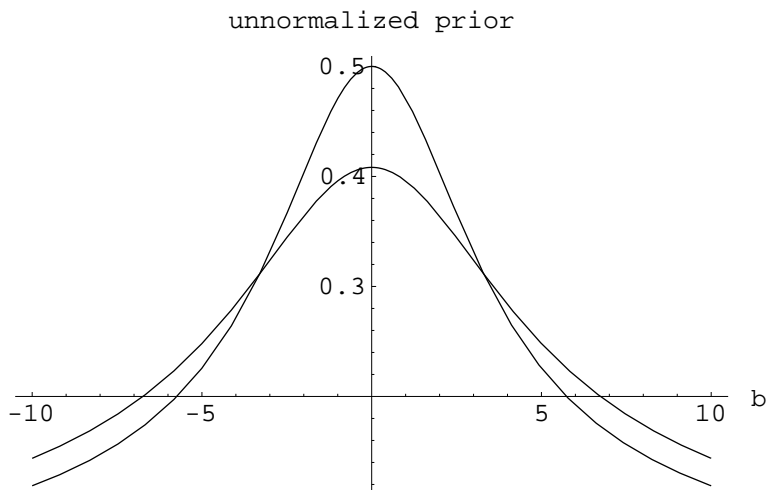}}
\caption{Unnormalized Jeffreys' priors over inverse temperature parameter
$\beta$. The lower-peaked curve corresponds to the quaternionic ($n=2$) case.}
\label{f6}
\end{figure}

Since for $n=1,2$, in particular, the Gibbs distributions (\ref{12}) form two
families of probability distributions, each parameterized by $\beta$, one can
find the (classical) Fisher information associated with them
 \cite{lav1,lav2,bern}. This is accomplished by computing the negatives of the expected values
(relative to (\ref{12})) of the second derivatives of the logarithms of
(\ref{12}) ($n=1,2$) with respect to $\beta$. The required integrations were
performed numerically. The square roots of the obtained Fisher information
statistics are exhibited in Fig.~\ref{f6} for $\beta \in [-10,10]$, with
the quaternionic ($n=2$) case having the smaller maximum.
 These serve as the
Jeffrey's (unnormalized) priors over the inverse temperature parameter
($\beta$), in the sense of Lavenda's pioneering investigations
 \cite{lav1,lav2,lav3}.
 (Fig.~\ref{f6} was also reproducible, with MATHEMATICA, using
exact/numerical methods, but only by taking the required integrations over
$z \in [-1,1]$ --- which themselves proved to be problematical --- to be the
results of integrations over $z \in [0,1]$ and the addition to them of the
outcomes of substituting $-\beta$ for $\beta$ in them.)

For literature supplemental to \cite{lav1,lav2} pertaining to the topic of
temperature fluctuations, see
\cite{tiko,szil,mand1,mand2,chak,chui,pros,slat3,stod}.
The quantum Fisher information metric \cite{fuji} employed here
to generate the structure (prior) functions (\ref{8}) and (\ref{9}) has been
shown \cite{brau} to be simply proportional to the Bures metric --- which
extends to the mixed states, the Fubini-Study metric on the pure states
of a quantum system.

\acknowledgments

I would like to express appreciation to the Institute for Theoretical
Physics for computational support in this research.


\begin{references}
\bibitem{lav1} Lavenda B. H., Int. J. Theor. Phys. 27 (1988) 451.
\bibitem{lav2} Lavenda B. H., Statistical Physics: An Introduction
(Wiley, New York) 1991.
\bibitem{lav3} Lavenda B. H., Int. J. Theor. Phys. 34 (1995) 615.
\bibitem{adler} Adler S. L., Quaternionic Quantum Mechanics and Quantum
Fields (New York, Oxford) 1995.
\bibitem{belt} Beltrametti E. G. and Cassinelli G., The Logic of Quantum
Mechanics (Reading, Addison-Wesley) 1981.
\bibitem{hels} Helstrom C. W., Quantum Detection and Estimation Theory
(New York, Academic) 1976.
\bibitem{khol} Kholevo A. S., Probabilistic and Statistical Aspects of
Quantum Theory (Amsterdam, North-Holland) 1982.
\bibitem{mall} Malley J. D. and Hornstein J., Statist. Sci. 8 (1993) 433.
\bibitem{fuji} Fujiwara A. and Nagaoka H., Phys. Lett. A 201 (1995) 119.
\bibitem{brau} Braunstein S. L. and Caves C. M., Phys. Rev. Lett.
72 (1994) 3439.
\bibitem{slat} Slater P. B., J. Math. Phys.  37 (1996) 2682.
\bibitem{bied} Biedenharn L. C. and Louck J. D., Angular Momentum in
Quantum Physics (Reading, Addison-Wesley) 1981.
\bibitem{jeff1} Jeffreys H., Proc. Roy. Soc. Lond. A 186 (1946) 453.
\bibitem{jeff2} Jeffreys H., The Theory of Probability (Oxford, Clarendon) 
1961.
\bibitem{bern} Bernardo J. M. and Smith A. F. M., Bayesian Theory (New York,
Wiley) 1994.
\bibitem{karm} Karmazina L. N. and Prudnikov A. P., in Encyclopaedia of
Mathematics, eidted By M. Hazewinkel (Dordrecht, Kluwer) 1988, vol. 2, pp.
54-507.
\bibitem{park} Park J. L. and Band W., Found. Phys. 7 (1977) 233.
\bibitem{band} Band W. and Park J. L., Found. Phys. 7 (1977) 705.
\bibitem{slat2} Slater P. B., Phys. Lett. A 171 (1992) 285.
\bibitem{tiko} Tikochinsky Y. and Levine R. D., J. Math. Phys. 25 (1984) 2160.
\bibitem{szil} Szilard L., Z. Phys. 32 (1925) 753.
\bibitem{mand1} Mandelbrot B. B., J. Math. Phys. 5 (1964) 164.
\bibitem{mand2} Mandelbrot B. B., Phys. Today 42:1 (1989) 71.
\bibitem{chak} Chakrabarti C. G., Czech. J. Phys. B 29 (1979) 837, 841.
\bibitem{chui} Chui T. R. {\it et al}, Phys. Rev. Lett. 69 (1992) 3005.
\bibitem{pros} Prosper H. B., Amer. J. Phys. 61 (1993) 54.
\bibitem{slat3} Slater P. B., Phys. Lett. A 176 (1993) 184.
\bibitem{stod} Stodolsky L., Phys. Rev. Lett. 75 (1995) 1044.

\end{references}
\end{document}